\documentclass[fleqn,usenatbib]{mnras}

\usepackage[dvipdfmx]{graphicx}

\usepackage{amsmath}
\usepackage{color}

\usepackage[T1]{fontenc}
\usepackage{aecompl}
\usepackage{yfonts}

\begin{document}

\title[]{Kinematic fingerprint of core-collapsed globular clusters}

 \author[P. Bianchini et al.]{P. Bianchini$^{1,}$\thanks{E-mail:
bianchip@mcmaster.ca}\thanks{CITA National Fellow},
J. J. Webb$^{2,3}$,
A. Sills$^{1}$,
E. Vesperini$^{3}$
\\
$^{1}$Department of Physics and Astronomy, McMaster University, Hamilton, ON, L8S 4M1, Canada\\
$^{2}$Department of Astronomy and Astrophysics, University of Toronto, 50 George Street, Toronto, ON, M5S 3H4, Canada\\
$^{3}$Department of Astronomy, Indiana University, Swain West, 727 E. 3rd Street, IN 47405 Bloomington, USA
}

\date{Accepted 2018 January 22. Received 2018 January 21; in original form 2017 November 27}
\maketitle

\begin{abstract}
Dynamical evolution drives globular clusters toward core collapse, which strongly shapes their internal properties. Diagnostics of core collapse have so far been based on photometry only, namely on the study of the concentration of the density profiles. Here we present a new method to robustly identify core-collapsed clusters based on the study of their stellar kinematics. We introduce the \textit{kinematic concentration} parameter, $c_k$,  the ratio between the global and local degree of energy equipartition reached by a cluster, and show through extensive direct $N$-body simulations that clusters approaching core collapse and in the post-core collapse phase are strictly characterized by $c_k>1$. The kinematic concentration provides a suitable diagnostic to identify core-collapsed clusters, independent from any other previous methods based on photometry. We also explore the effects of incomplete radial and stellar mass coverage on the calculation of $c_k$ and find that our method can be applied to state-of-art kinematic datasets.
\end{abstract}



\section{Introduction}
Globular clusters (GCs) are old stellar systems significantly shaped by the internal two-body gravitational interactions between their stars. A major outcome of GC secular evolution is the gravothermal instability, leading to the collapse of the core (e.g. \citealp{Henon1961,Lynden-BellWood1968}). The process of core collapse strongly affects their structural, morphological and kinematic properties.

A common way to observationally classify clusters that have undergone core collapse relies on the study of their surface density profiles. Core-collapsed clusters are identified as those with density profiles exhibiting a central cusp described by a power-law, as opposed to pre-core collapsed clusters characterized by profiles following a simple one-component King model with a flat core (e.g. \citealp{DjorgovskiKing1986,Chernoff1989,Trager1995}). This distinction can, however, not be definite as clusters with a central photometric cusp can also be successfully described by high-concentration King models or multi-mass models (see \citealp{MeylanHeggie1997} and reference therein).

According to this classification approximately 20\% of the Milky Way (MW) GCs are core collapsed. However, there is no robust connection between their central concentration and their actual dynamical state. In fact, physical processes such as binary heating can induce a reexpansion of the core and significantly complicate the interpretation of the current state of a cluster based solely on the study of its density profile (\citealp{Chernoff1989,Heggie2009}).  

In this Letter, we propose a new and complementary approach to identify core-collapsed clusters based solely on internal kinematics. Our approach consists of studying the evolution of energy equipartition in a GC, driven by the redistribution of energy between stars through their mutual gravitational interactions. While massive stars lose energy and sink toward the central regions, low-mass stars gain energy and move outwards. This results in a mass segregated cluster, characterized by a mass-dependent velocity dispersion (e.g. \citealp{WebbVesperini2017,Bianchini2016b,TrentivanderMarel2013}). Measurements of the degree of energy equipartition are now within reach of observations thanks to state-of-the-art \textit{Hubble Space Telescope} (\textit{HST}) proper motion datasets (e.g. \citealp{Heyl2017,Bellini2018} and HSTPROMO collaboration, \citealp{Bellini2014}) or Gaia proper motions \citep{Pancino2017}, allowing us to measure the velocity dispersion profile as a function of stellar mass.

Here we study the specific relation between the global and local degree of energy equipartition and the process of core collapse of a GC, using an extensive set of $N$-body simulations (Section \ref{sec:2}). From our models we are able to identify a kinematic fingerprint of core collapse using the newly introduced \textit{kinematic concentration} parameter (Section \ref{sec:3} and \ref{sec:4}).



\section{Simulations}
\label{sec:2}

\begin{figure}
\centering
\includegraphics[width=0.48\textwidth]{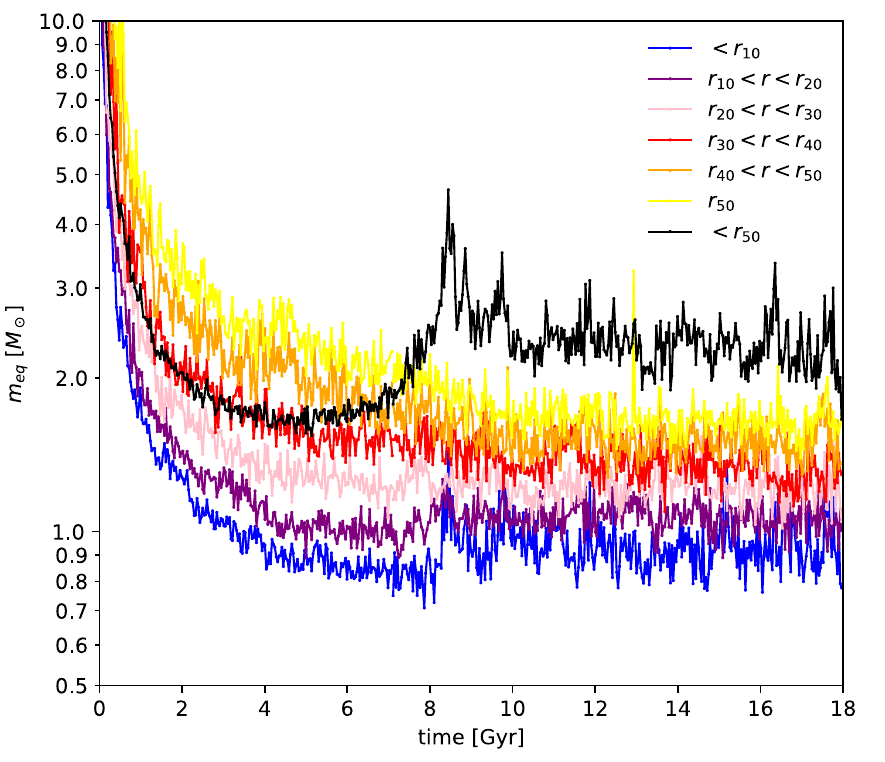}
\caption{Time evolution of the energy equipartition parameter $m_{eq}$ (see eq. 1) measured at different Lagrangian radii for simulation MW20. From bottom to top: within 10\% Lagrangian radius (blue line), 10-20\% (purple line), 20-30\% (pink line), 30-40\% (red line), 40-50\% (orange line), at 50\% Lagrangian radius (yellow line) and within the 50\% Lagrangian radius (black line). A radial dependence of $m_{eq}$ is observed, with the central regions of the cluster reaching a state closer to full equipartition (i.e. lower $m_{eq}$). The $m_{eq}$ within $r_{50}$ (black line) exhibits a non-monotonic behaviour; this is due to the interplay of mass segregation, core collapse and the radial-dependent velocity dispersion.}
\label{fig:meq_timeev}
\end{figure}

We collect a sample of 26 direct $N$-body simulations originally presented in \citet{Bianchini2017b} (see also \citealp{Miholics2016}) and \citet{WebbVesperini2016a}, using the $N$-body codes Nbody6tt \citep{Renaud2011,Renaud2015} and  Nbody6 \citep{Aarseth2003,Nitadori2012}. The simulations from  \citealp{Bianchini2017b} include a \citet{Kroupa2001} initial mass function, while the ones from \citealp{WebbVesperini2016a} utilize a \citet{Kroupa1993} mass function. All the simulations have lower and upper stellar mass limits of 0.1 and 50 $M_{\odot}$, include stellar evolution and the presence of a tidal field. The initial configurations are sampled from a \cite{Plummer1911} distribution and encompass a large range of initial configurations (number of particles N=$50,000-100,000$ and initial half-mass radius $r_{50}=1.1-6$ pc; see Table \ref{tab:initial}). Simulations from \citet{Bianchini2017b} are either evolved in a variety of time-dependent tidal environments describing the evolution of a GC accreting onto the MW from a disrupting dwarf galaxies, or in circular orbit around the MW or a dwarf galaxy (see \citealp{Bianchini2017b} and \citealp{Miholics2016} for details). 
Simulations from \citet{WebbVesperini2016a} include clusters with eccentric orbits, primordial binaries, and/or non-zero black hole retention fractions.

\begin{table}
\tabcolsep=0.10cm
\begin{center}
\caption{Initial conditions of our set of simulations originally presented in (1) \citet{Bianchini2017b} and (2) \citet{WebbVesperini2016a}. Here we report the initial number of particles $N$, the initial half-mass radius $r_{50}$, the distance to the galactic centre $d_{GC}$ and the mass of the host dwarf galaxy $M_{DW}$, if present. The primordial binary fraction, black holes retention fraction and eccentricity of the orbit are specified; if nothing is specified, the percentage of primordial binaries and retained black holes is zero, and the orbits are circular. The detailed description of the simulations set up is presented in the corresponding papers.}

\begin{tabular}{lcccccc}
\hline\hline
 &$N$ & $r_{50}$ & $d_{GC}$ & $M_{DW}$ & comments&  ref.\\

\hline
MW only & $\times 10^3$& pc & kpc & $M_\odot$& &\\
\hline
MW10	& 50	&3.2 & 10& $-$& $-$& (1)\\
MW15	& 50	&3.2 & 15& $-$& $-$& (1)\\
MW15-R1.6& 50	&1.6 & 15& $-$& $-$& (1) \\
MW15-R4& 50	&4.0 & 15& $-$& $-$& (1)\\
MW20	& 50	&3.2& 20& $-$& $-$& (1)\\
MW30	& 50	&3.2 & 30& $-$& $-$& (1) \\
E0R6RM1 & 100& 1.1 & 6 & $-$& $-$&(2)\\
E0R6RM1B4 &100 &  1.1 & 6 &$-$ & $4\%$ bin&(2)\\
E05RP6B4 &100 &  1.1 & 6 &$-$ & $4\%$ bin, $e$=0.5&(2)\\
E0R18B4 &100 &  1.1 & 18 & $-$& $4\%$ bin&(2)\\
E0R6RM1BH25 &100 &  1.1 & 6 &$-$ & $25\%$ BHs&(2)\\
E0R6RM1BH50 &100 &  1.1 & 6 &$-$ & $50\%$ BHs&(2)\\
\hline
Dwarf only&&&&\\
\hline
DWL& 	50	&3.2& 4& $10^{10}$& $-$& (1)\\
DWS& 	50	&3.2&  4& $10^{9}$& $-$& (1)\\
DWS-R1.6& 50	&1.6& 4& $10^{9}$& $-$& (1)\\
DWS-R4&	 50	&4.0& 4& $10^{9}$& $-$& (1)\\

\hline
Accreted& &&& \\

\hline
DWL-MW10-evap&	50	&3.2 & 10& $10^{10}$ & $-$& (1)\\
DWS-MW10-evap&	50	&3.2 & 10& $10^{9}$& $-$& (1)\\
DWL-MW20-evap&	50	&3.2 & 20& $10^{10}$& $-$& (1)\\

DWL-MW10-fal&	50	&3.2 & 50& $10^{10}$& $-$& (1)\\
DWS-MW10-fal&	50	&3.2 & 50& $10^{9}$& $-$& (1)\\
DWL-MW15-fal&	50	&3.2 & 50& $10^{10}$& $-$& (1)\\
DWS-MW15-fal&	50	&3.2 & 50& $10^{9}$& $-$& (1)\\
DWS-MW15-R1.6-fal&  50 	&1.6& 50& $10^{9}$& $-$& (1)\\
DWS-MW15-R4-fal&	 50	&4.0& 50& $10^{9}$&  $-$&(1)\\
DWS-MW30-fal&	 50	&3.2& 50& $10^{9}$& $-$& (1)\\

\hline

\end{tabular}
\label{tab:initial}
\end{center}
\end{table}

\begin{figure*}
\centering
\includegraphics[width=0.85\textwidth]{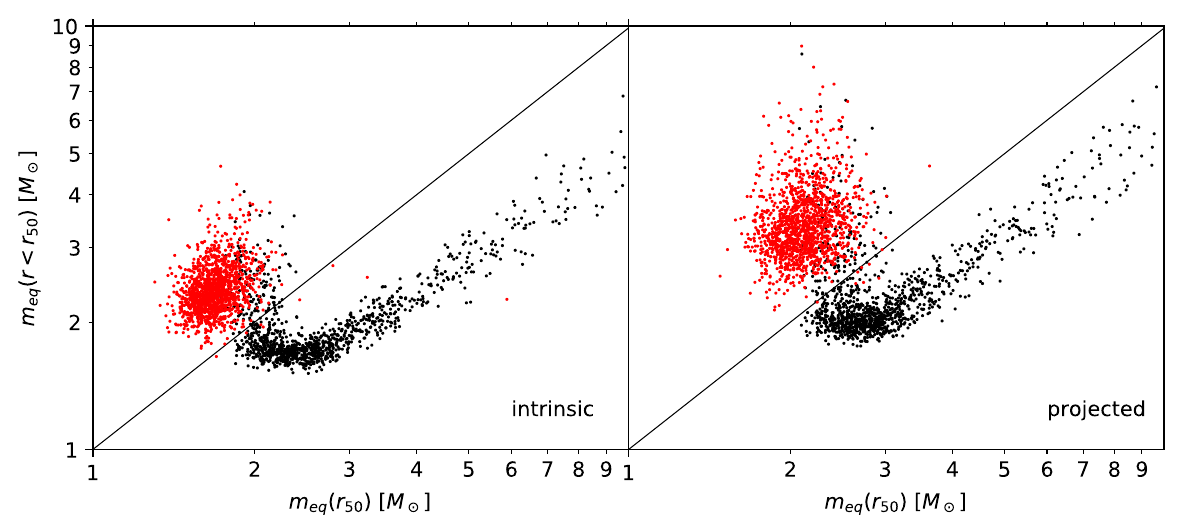}
\caption{Level of energy equipartition within the half-mass radius $m_{eq}(r<r_{50})$ versus the level of energy equipartition at the half-mss radius $m_{eq}(r_{50})$, for simulation MW20. \textit{Left panel:} Intrinsic $m_{eq}$ measured from the velocity dispersions as a function of the intrinsic radius. \textit{Right panel:} $m_{eq}$ measured from the projected snapshot. Red points indicate post-core collapse snapshots and lie above the 1:1 line (the few exceptions are related to the cluster being close to dissolution). This indicates that post-core collapse clusters exhibit a value of $m_{eq}(r<r_{50})$ higher than the local $m_{eq}(r_{50})$, particularly evident when measured in projection (right panel). This can be explained by the effect of strong mass segregation occurring during core collapse, affecting mainly the inner regions of the cluster.}
\label{fig:meq_global_local}
\end{figure*}
\begin{figure*}
\centering
\includegraphics[width=0.66\textwidth]{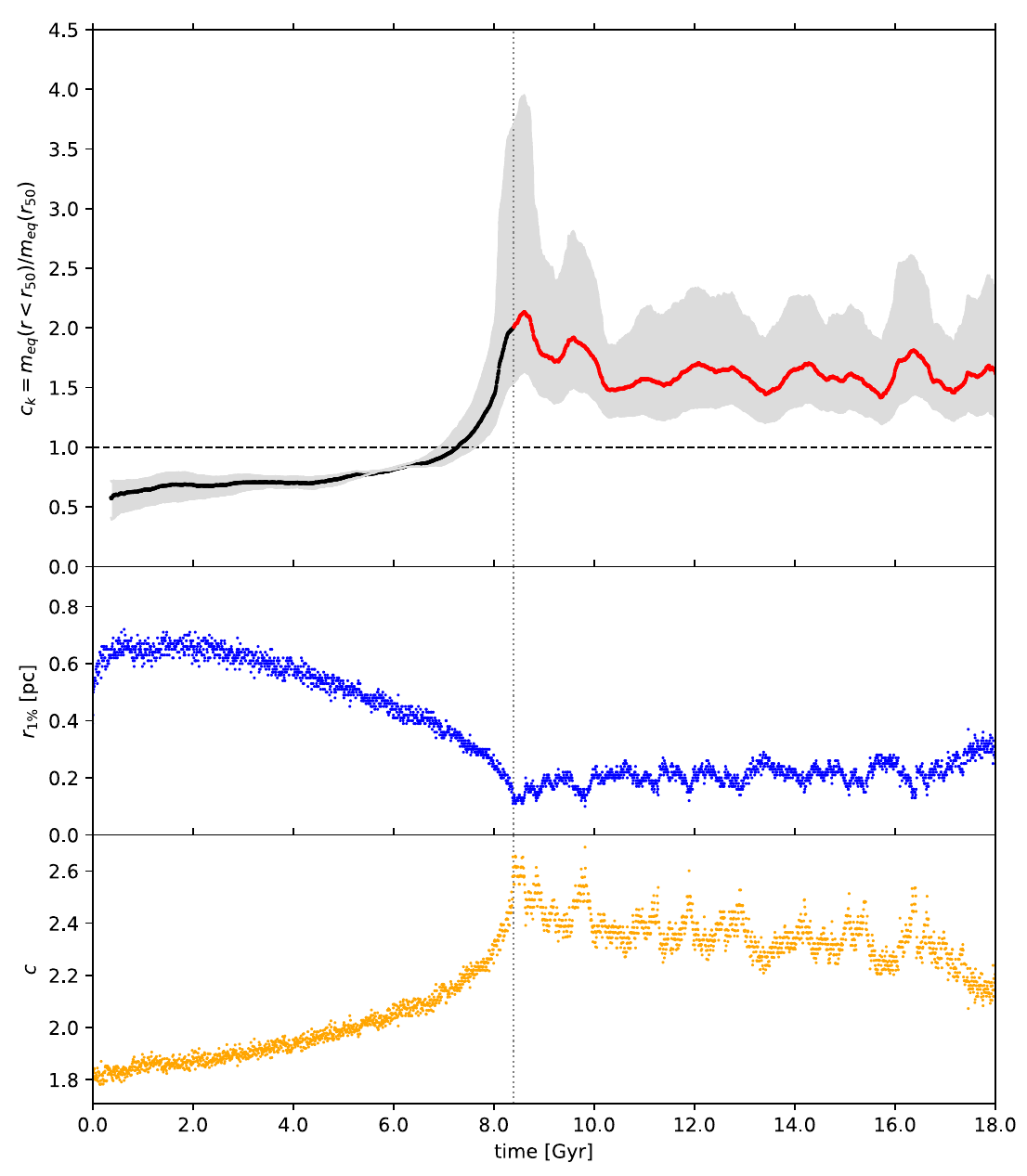}
\caption{\textit{Top panel:} Time evolution of the kinematic concentration $c_k=m_{eq}(<r_{50})/m_{eq}(r_{50})$ for the projected snapshots of simulation MW20. The red line indicates snapshots after core collapse, the shaded area indicates values of $c_k$ measured using the 40\% and 60\% Lagrangian radii (lower and upper limit, respectively). The curves have been smoothed by averaging the values of $c_k$ of 100 consecutive snapshots. \textit{Central panel:} Time evolution of the 1\% Lagrangian radius, $r_{1}$, indicating the collapse of the core at $\simeq8.4$ Gyr. \textit{Bottom panel:} Time evolution of the photometric concentration $c=\log(r_{99}/r_1)$. When the cluster approaches core collapse, and throughout the post-core collapse phase, the value of kinematic concentration $c_k$ reaches values $>1$. This is a strong kinematic signature of core collapse.}
\label{fig:corecollapse_meq}
\end{figure*}

\subsection{Measuring the degree of energy equipartition: local vs. global}
\label{sec:3}
For each time step in our simulations we compute the degree of energy equipartition reached by the cluster using the method introduced by \citet{Bianchini2016b} based on a fit of an exponential function to the mass-dependent velocity dispersion $\sigma=\sigma(m)$
\begin{equation}
\sigma(m)=\sigma_0 \exp\left(-\frac{1}{2}\frac{m}{m_{eq}}\right ),
\label{eq:meq}
\end{equation}
with $\sigma_0$ indicating the extrapolated value of the velocity dispersion at $m=0$, and the parameter $m_{eq}$ the degree of partial energy equipartition reached by the system: \textit{clusters characterized by lower values of $m_{eq}$ are closer to full energy equipartition.}
The single parameter $m_{eq}$ uniquely describes the shape of $\sigma(m)$ at lower as well as at higher stellar masses, and provides a more flexible description of $\sigma=\sigma(m)$ than the traditionally used power-law $\sigma\propto m^{-\eta}$ (e.g. \citealp{TrentivanderMarel2013}).\footnote{The equipartition parameter $m_{eq}$ can be converted to a local $\eta$ parameter using eq. 4 of \citet{Bianchini2016b}.}

The mass-dependent velocity dispersion curves are calculated considering the following stellar mass bins: $m<0.3$, $0.3<m<0.4$, $0.4<m<0.5$, $0.5<m<0.7$, $m>0.7$~$M_\odot$. Since the level of equipartition reached by a cluster is a radially dependent quantity (because it directly depends on the relaxation time of a cluster, which is a local quantity), we compute $\sigma(m)$ locally within different Lagrangian radii\footnote{The Lagrangian radii are calculated as the radii containing a given percentage of the total mass.} as well as globally within the 50\% Lagrangian radius. The velocity dispersion is calculated as the average of the three velocity dispersion components, $\sigma^2=(\sigma_x^2+\sigma_y^2+\sigma_z^2)/3$. For each of the local and global $\sigma(m)$ curves we estimate the degree of equipartition with the parameter $m_{eq}$.

Figure \ref{fig:meq_timeev} shows the time evolution of $m_{eq}$ for one of the simulations (MW20, GCs evolving in a circular orbit at 20 kpc around the MW's centre) calculated locally within different Lagrangian radii as well as globally within $r_{50}$. While dynamical evolution proceeds, the system approaches a state closer to full equipartition (decreasing $m_{eq}$). A clear radial dependence of $m_{eq}$ is observed, with the central regions of the cluster reaching a state closer to full equipartition, in accordance with the shorter relaxation times in these regions. When core collapse is approached ($\simeq8$ Gyr), $m_{eq}$ in the inner Lagrangian radius increases, indicating stronger interaction between massive stars in the central region due to the collapse of the core, enhancing their velocity dispersion. After core collapse, at every Lagrangian radii, the process of equipartition significantly slows down (in agreement with \citealp{Webb2017,GierszHeggie1996})

The global value of $m_{eq}$ within $r_{50}$ exhibits a non-monotonic behaviour, with a peak at $\simeq8$ Gyr. This behaviour is due to the interplay between the radial dependence of the velocity dispersion and mass segregation. Since high-mass stars fall in towards the cluster centre, the high-mass end of $\sigma(m)$ will become dominated by stars with a higher velocity dispersion as the cluster undergoes mass segregation. Conversely, with lower mass stars migrating outwards, the low-mass end of $\sigma(m)$ will become dominated by stars with a lower velocity dispersion since $\sigma$ decreases with clustercentric distance \citep{WebbVesperini2017}. Since the central velocity dispersion will continue to increase as the cluster approaches core collapse, $m_{eq}$ within $r_{50}$ will stop decreasing and start to increase, despite $m_{eq}$ continuing to decrease locally throughout the cluster.

In the rest of our work we will consider the relation between the evolution of the local and global $m_{eq}$ parameters ($m_{eq}(r_{50})$ and $m_{eq}(r<r_{50})$) and the onset of core collapse. For this purpose we introduce a new parameter named the \textit{kinematic concentration}, $c_k$, which is analogous to the standard photometric concentration parameter $c$ (e.g. \citealp{Harris1996}, 2010 edition) and will be used as an indicator of core collapse. \textit{The kinematic concentration is defined as the ratio between the degree of equipartition reached globally within the half-mass radius and locally at the half-mass radius,}
\begin{equation}
c_k=\frac{m_{eq}(r<r_{50})}{m_{eq}(r_{50})}.
\label{eq:c_k}
\end{equation}

The local $m_{eq}(r_{50})$ is calculated considering stars between the 40\% and 60\% Lagrangian radii, $r_{40}$ and $r_{60}$. All the following figures will refer to simulation MW20, but equivalent results hold for our entire set of simulations. 
Moreover, to facilitate a comparison with observations, we will measure $c_k$ using both the intrinsic and projected properties of stars in the cluster, as specified in rest of the text.

\section{Kinematic fingerprint of core collapse}
\label{sec:4}
In Figure \ref{fig:meq_global_local} we plot the global measure of equipartition $m_{eq}(r<r_{50})$ versus the local measure of $m_{eq}(r_{50})$ for all the time snapshots of our reference simulation. In the left panel, we show the results for the intrinsic quantities, while in the right panel we consider snapshots projected on to the plane of the sky. In both cases, the snapshots start away from equipartition (both globally and locally, i.e. high values of $m_{eq}$, top right corner of each plot) and progressively approach higher degrees of energy equipartition, until reaching a minimum at around $m_{eq}(r_{50})\sim2-3$ $M_\odot$. Until this point the global value of $m_{eq}$ is always lower than the local value. Once the cluster reaches core collapse (8.4 Gyr, red points), the global value of $m_{eq}$ increases and settles to values higher than the local $m_{eq}$. In the post-core collapse phase, the global $m_{eq}$ remains higher than the local value. This result is stronger for the projected snapshots.

\begin{figure*}
\centering
\includegraphics[width=0.482\textwidth]{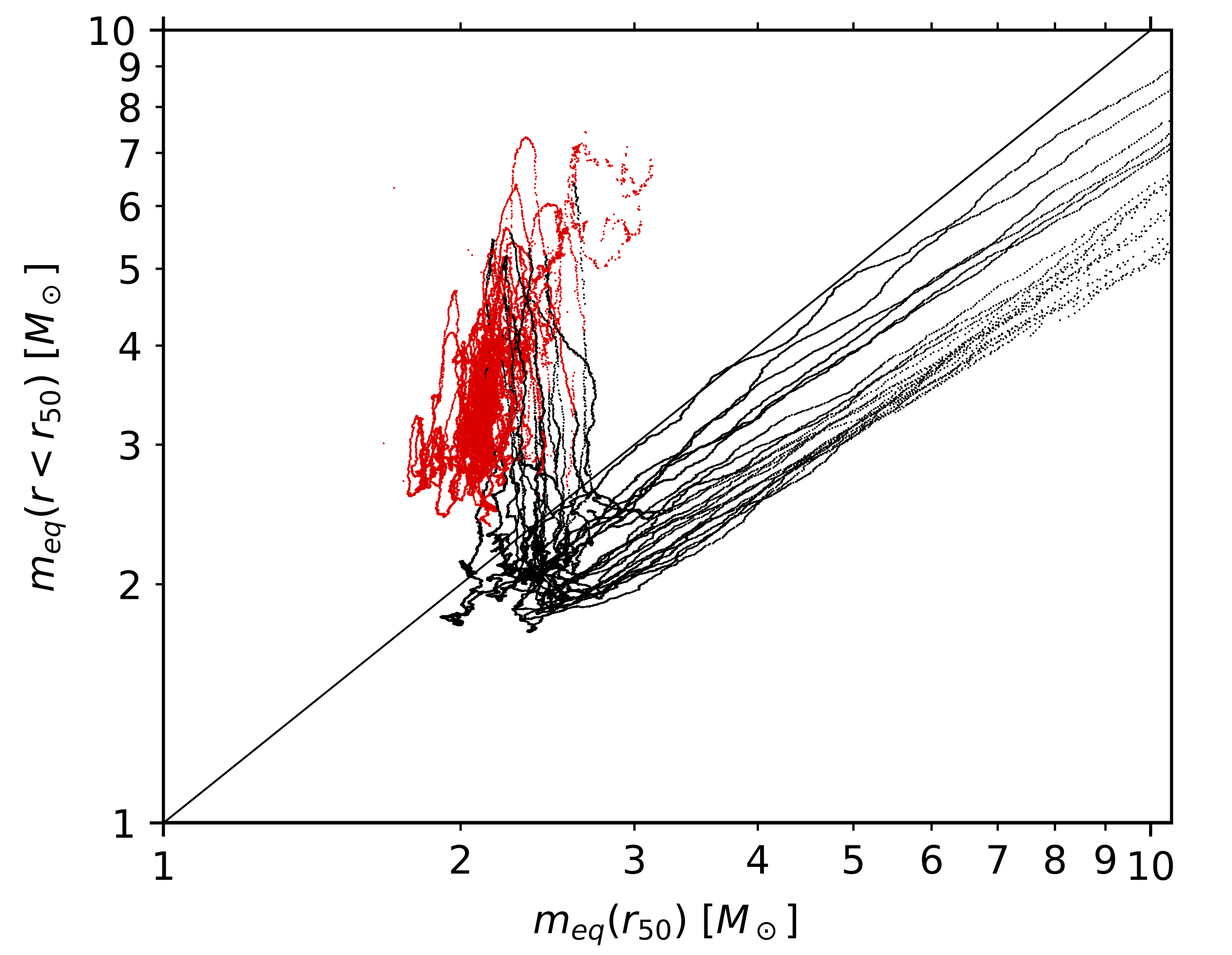}\,
\includegraphics[width=0.48\textwidth]{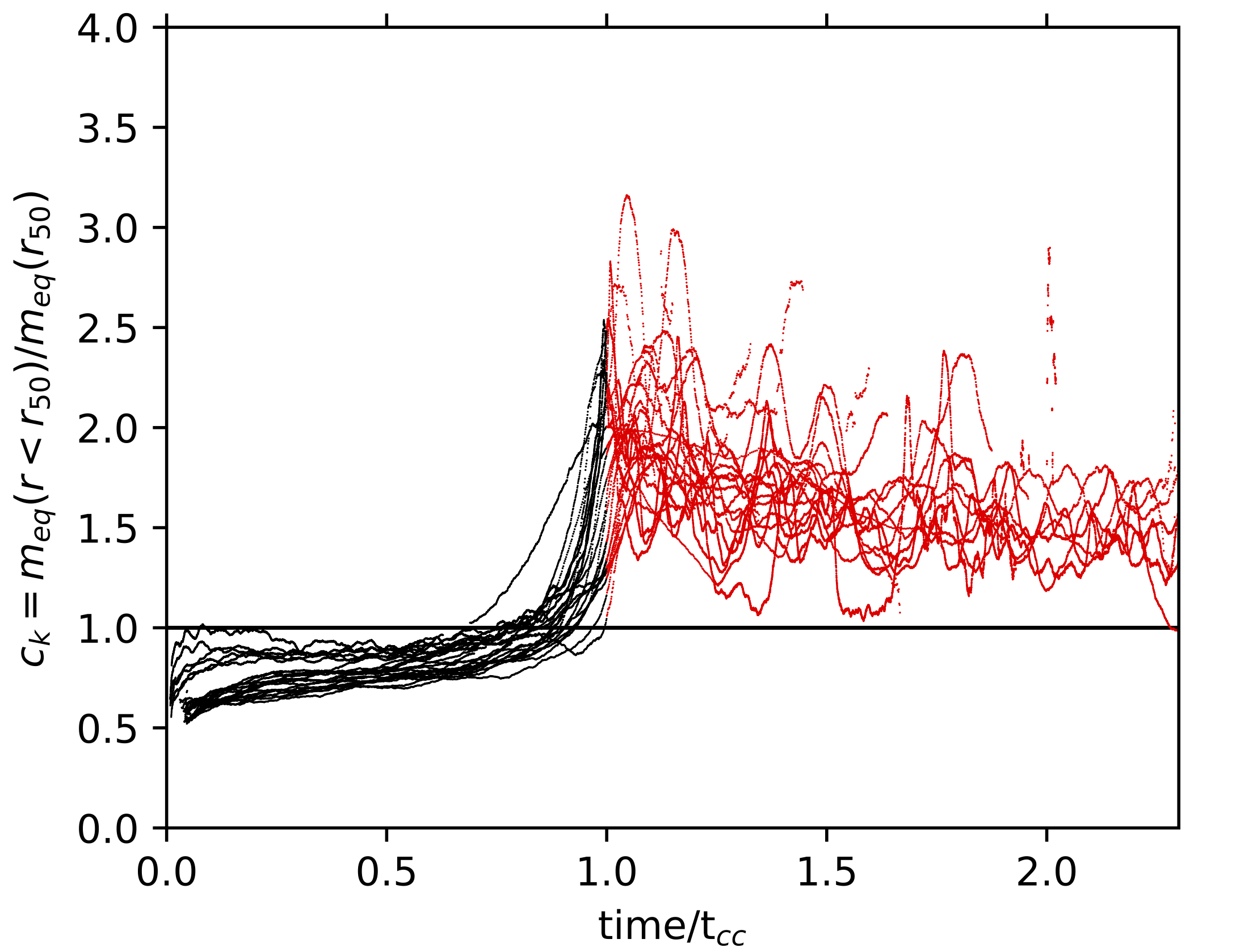}
\caption{\textit{Left panel:} Level of energy equipartition within the half-mss radius $m_{eq}(r<r_{50})$ versus the level of energy equipartition at the half-mss radius $m_{eq}(r_{50})$ for all our simulations. \textit{Right panel:}  Evolution of the kinematic concentration $c_k=m_{eq}(<r_{50})/m_{eq}(r_{50})$ as a function of time (normalized by the time of core collapse $t_{cc}$), for all our simulations. Red points indicate snapshots in the post-core collapse phase and they are all characterized by kinematic concentration $c_k>1$. All the snapshots are considered in projections, and the curves have been smoothed by averaging the values of $c_k$ of 100 consecutive snapshots.}
\label{fig:ALL}
\end{figure*}

In Figure \ref{fig:corecollapse_meq} we present the time evolution of the kinematic concentration $c_k$ (see eq. \ref{eq:c_k}) compared to the evolution of the 1\% Lagrangian radius $r_{1}$ (i.e. a proxy for the King core radius) and of the photometric concentration $c=\log(r_{99}/r_{1})$, with $r_{99}$ the 99\% Lagrangian radius, used as a proxy for the limiting radius.
All of these quantities are measured in projection. The plot clearly shows the correspondence between the peak of $c_k$ and the collapse of the core, indicated both by the minimum of $r_1$ and by the peak of $c$. 
The clusters reach values of kinematic concentration $c_k>1$ in the latest phases of core-collapse and maintain such values throughout the post-core collapse phase.
To explore how measurements of $c_k$ depend on the radial range used, we illustrate with the shaded region in the top panel of Figure \ref{fig:corecollapse_meq} the effect of measuring the local and global $m_{eq}$ using the 40\% and 60\% Lagrangian radii, $r_{40}$ and $r_{60}$, instead of $r_{50}$. Our result does not strongly depend on the choice of the Lagrangian radius.

It should be noted that the photometric concentration $c$ tends to decrease in the later phases of post-core collapse, yielding values that do not uniquely identify the cluster as being in the post-core collapse phase. $c_k$, on the other hand, stays above 1 indefinitely once core collapse has occurred. This suggests that a cluster will maintain values of $c_k>1$ even when reexpanding because of binary heating or presence of black holes.

In Figure \ref{fig:ALL} we show the evolution of the local and global $m_{eq}$ for all our simulations, indicating the validity of our results for the variety of initial configurations and dynamical histories described by our clusters.
Figures \ref{fig:corecollapse_meq} and \ref{fig:ALL} demonstrate that the kinematic concentration $c_k$ provides a robust kinematic signature of core collapse, completely independent of the standard photometric definition, and clearly connected to the physical processes taking place. We emphasize that the initial clusters' density, their black holes retention fraction, the presence of primordial binaries and the variety of tidal environments do not compromise the $c_k>1$ criterion for core collapse; however, these ingredients do affect the detail of the evolution of the $m_{eq}$ parameter (Fig. \ref{fig:ALL}, left panel). We defer the analysis of these dependencies to a subsequent investigation.
We note that while $c_k$ does technically become greater than 1 before core collapse occurs, this pre-core collapse phase lasts on average between $0.4-2.0$ Gyr (corresponding to $0.1-1.5$ half-mass relaxation times) such that it can be said with confidence that all clusters with $c_k>1$ are in the post core collapse phase, or close to it.

Finally, in view of an application to proper motion data sets, we have explored the robustness of our results after introducing a radial and mass dependent incompleteness to the simulation outputs.
Local measurements of the velocity dispersion at a given mass bin are not affected by incompleteness; however, global values could be biased, if no correction is taken into consideration (e.g. weighing the radial velocity dispersion profile with the number density). Our tests shows that even if the data are incomplete, the measure of the kinematic concentration $c_k$ can be recovered unbiased for both the pre-  and post-core collapse snapshots.

\section{Conclusions}
In this Letter we present a new diagnostic for identifying GCs that have undergone core collapse, independent from previous methods that are based on studying GC density profiles. Our method is based on stellar kinematics and the interplay between mass segregation/energy equipartition and core collapse.

We introduce the \textit{kinematic concentration} parameter, $c_k$, the ratio between the global and local degree of energy equipartition reached by a cluster, and show that its evolution traces the onset of core collapse. In particular, we demonstrate that values of kinematic concentration $c_k>1$ correspond to evolutionary phases concurrent with core collapse and to the post-core collapse phases. This result is independent from the variety of initial conditions and dynamical evolutions of our set of simulations.

We test the validity of our results on our simulations including data incompleteness in anticipation of direct application to observational data. This test confirms that the measure of the kinematic concentration $c_k$ is accessible through the study of the velocity dispersion profile as a function of the stellar mass, currently available for state-of-the-art proper motion datasets of a sample of MW GCs (see e.g. \citealp{Heyl2017} and \citealp{Bellini2018}).



\section*{Acknowledgments}
We would like to thanks Roeland van der Marel and Andrea Bellini for useful discussions. PB acknowledges financial support from a CITA National Fellowship.

\bibliographystyle{mnras} 
\bibliography{biblio} 

\end{document}